\documentclass[iop]{emulateapj-rtx4} % read by: ACROREAD
\shortauthors{Sekanina}
\shorttitle{1I/`Oumuamua As Debris of Interstellar Comet}
\slugcomment{Version \today }

\begin{document}
\title{1I/`OUMUAMUA AS DEBRIS OF DWARF INTERSTELLAR COMET\\THAT DISINTEGRATED
  BEFORE PERIHELION}
\author{Zdenek Sekanina}
\affil{Jet Propulsion Laboratory, California Institute of Technology,
  4800 Oak Grove Drive, Pasadena, CA 91109, U.S.A.}
\email{Zdenek.Sekanina@jpl.nasa.gov{\vspace{0.15cm}}}

\begin{abstract}
Intrinsically faint comets in nearly-parabolic orbits with perihelion distances
much smaller~than~1~AU exhibit strong propensity for suddenly disintegrating
at a time not long before perihelion, as shown by Bortle (1991).  Evidence
from available observations of such comets suggests that the disintegration
event usually begins with an outburst and that the debris is typically a massive
cloud of dust grains that survives over a limited period of time.  Recent CCD
observations revealed, however, that also surviving could be a sizable fragment,
resembling a devolatilized aggregate of loosely-bound dust grains that may have
exotic shape, peculiar rotational properties, and extremely high porosity, all
acquired in the course of the disintegration event.  Given that the brightness
of 1I/`Oumuamua's parent~could~not possibly equal or exceed the Bortle survival
limit, there are reasons to believe that~it~suffered the same fate as do the
frail comets.  The post-perihelion observations then do not refer to the object
that was entering the inner Solar System in early 2017, as is tacitly
assumed,~but~to~its~debris.~\mbox{Comparison}~with C/2017~S3 and C/2010~X1
suggests that, as a monstrous fluffy dust aggregate released in the recent
explosive event, `Oumuamua should be of strongly irregular shape, tumbling, {\it
not\/} outgassing,~and~subjected to effects of solar radiation pressure, consistent
with observation.  The unknown timing of the disintegration event may compromise
studies of the parent's home stellar system.  Limited search for possible images
of the object to constrain the time of the (probably minor) outburst is recommended.
\end{abstract}

\keywords{comets: individual (1I/`Oumuamua, 67P/Churyumov-Gerasimenko, C/2010 X1,
C/2011~W3, C/2017 S3) --- methods: data analysis}

\section{Perihelion Survival Limit}
Investigations of 1I/`Oumuamua tacitly assume that it had not fundamentally
changed between the times of its entry to the inner Solar System (at $\sim$5~AU
from the Sun) in early 2017 and its discovery some nine months later.  While
there is no information, positive or negative, on any such potential changes,
I argue below that they were bound to happen, if the behavior of intrinsically
faint comets in nearly-parabolic orbits should apply to 1I.\footnote{The
alternative, asteroidal model for 1I was shown to be indefensible on other
grounds, especially because of the presence of a nongravitational acceleration
(Micheli et al.\ 2018).{\vspace{-0.3cm}}}

Thirty-five years ago I wrote a paper on several comets that had suddenly
begun to fade before perihelion, disappearing altogether soon afterwards
(Sekanina 1984).  They all were faint objects moving in mostly parabolic
orbits with perihelia below 1~AU.  Seven years later Bortle (1991) called
attention to the fact that virtually {\small \bf no intrinsically faint
long-period comets with perihelion distances smaller than about 0.25~AU
have ever been observed after perihelion} because they all perished
before, or in close proximity of, perihelion passage.  Bortle then
employed a set of long-period comets to determine that, at a confidence
level of 70~percent, there is a correlation between the perihelion
distance $q$ and the limiting total visual absolute magnitude $H_0(q)$
(normalized to 1~AU from both the Sun and Earth) that a comet should
have in order to survive perihelion passage essentially intact; fainter
comets disintegrate.  Since 1991 additional comets with small perihelion
distance have been observed to validate Bortle's rule.

\section{1I/`Oumuamua: The Object and Its Debris}
The survival limit for `Oumuamua's absolute magnitude at its perihelion
distance of 0.255~AU is \mbox{$H_0 = 8.5$}.  On the assumption that the
object brightened with heliocentric distance $r$ at the rate of an average
Oort Cloud comet, as $r^{-2.44}$ (Whipple 1978), its apparent magnitude
should have reached about 17.5 in mid-November 2016
($\sim$10~months before perihelion); approximately 16 in mid-February 2017
($\sim$7~months before perihelion); near 14.5 in mid-April ($\sim$5~months
before perihelion); close to 12 in mid-June ($\sim$3~months before perihelion);
and about 9 in mid-August ($\sim$1~month before perihelion).  And at perihelion
it could have been as bright as magnitude~5\,(!), more than 8~mag above the
threshold quoted by Bannister et al.\ (2017) with reference to the object's
nondetection by K.\ Battams in the STEREO-A HI1 images taken at the time.
`Oumuamua's path projected onto the plane of the sky was
peculiar in that the solar elongation was confined to between 47$^\circ$
and 106$^\circ$ over a period of one year, from 13~months to 1~month before
perihelion.

Although it is unknown how bright `Oumuamua was when approaching perihelion
from interstellar space, it certainly was {\small \bf much fainter than
required by Bortle's survival limit}.  An estimate is provided by its failure
to show up in the images taken with the Pan-STARRS1 telescope on June~17--22,
2017 (Micheli et al.\ 2018), when the object was at a heliocentric distance
of 2.18~AU.  Assuming the telescope's limiting magnitude of $\sim$21, the
absolute magnitude of `Oumuamua could not be brighter than $\sim$18,
suggesting that the object fits the category of {\it dwarf\/} comets.

Given now that `Oumuamua was missed~on the way to perihelion because of its
{\small \bf intrinsic faintness}, more than 9~mag below the survival limit,
it has since the very beginning been a {\small \bf perfect candidate for
disintegration} by virtue of its {\small \bf short perihelion distance\/}.  If
so, the post-perihelion observations refer not to the object that had entered
the inner Solar System, but to its {\small \bf debris}.  By investigating
the debris, one is not likely to learn very much about the object.  Many,
if not all, basic observed properties of 1I were probably acquired during the
disintegration event.  With less sophisticated detection methods of the
past, the debris would not have been discovered.  The state-of-the-art
instrumentation allows us to try and grasp the meaning of the terms
``disintegration'' and ``disappearance''.  

\section{Sudden Disintegration of Faint Comets\\Near the Sun}
Observations of comets show that a disintegration or fragmentation\footnote{By
{\it fragmentation\/} I understand here a crumbling of a comet's nucleus into
mostly very small pieces, not its {\it splitting\/} into two or a few more
nuclear components, even though both processes appear to be of similar nature.}
event of an {\it intrinsically faint comet\/} in a nearly-parabolic orbit
close to perihelion does not take much time and can easily be missed, in fact
with high probability so when it takes place before discovery.  Typically, the
event begins with an {\it outburst\/}, which opens up a narrow window of detection
opportunity.  Depending on the nucleus' dust-to-ice ratio, however, this window
may be as short as \mbox{2--3}~days.\footnote{Richter (1949) suggested that some
of the unconfirmed discoveries of the comets that were lost only a few days
later could be explained by their having been caught in outburst.}  First the
comet displays a star-like nuclear condensation, which rapidly expands
and becomes gradually more diffuse until it vanishes completely.  Although
the comet's {\it total\/} intrinsic brightness subsides more gradually,
by the time of perihelion passage the dimming usually becomes overwhelming and
in ground-based telescopes the object vanishes in the Sun's glare to never
re-emerge unscathed.

The ultimate product of disintegration that follows a cataclysmic outburst
is an expanding cloud of dust particles, minor fragments of the pulverized
nucleus.  They are subject to solar radiation pressure, so the
orbital motion of a disintegrated comet --- that is, of its debris ---
differs from the motion of the original comet.  The debris resulting from
a nonfatal outburst may be dominated by a sizable fragment (Section~4) that
may have no reservoir of ice, yet displays the capability to survive on
its own over a reasonably long period of time.  Because of its lower
area-to-mass ratio compared to dust grains, a sizable fragment rid of volatiles
is very hard to detect.

The process of sudden disintegration of faint comets is of course analogous
to the process of disintegration of the Kreutz system's dwarf sungrazers,
except that the physical conditions and scales are different.  The dwarf
sungrazers are subjected to much higher temperatures, {\it sublimating
away\/} entirely with no terminal outburst.

\section{Disintegration of Comet C/2017 S3}
An Oort Cloud comet C/2017 S3 (Pan-STARRS) provides an exceptional amount
of data on cometary disintegration and disappearance (Sekanina \& Kracht
2018), including information that is relevant to `Oumuamua.  The comet, of a
perihelion distance of 0.21~AU, had a total visual absolute magnitude
of 10.7 on its approach to perihelion, 2.5~mag fainter than Bortle's
survival limit.  Expected to disintegrate, the object experienced two
successive outbursts before perihelion, each having been accompanied by
a fragmentation event.  Evidence strongly suggests that the comet's
nucleus did indeed fall apart completely before reaching 0.9~AU from
the Sun.

Sekanina \& Kracht's results on the second, cataclysmic outburst indicate
that it began 32.6~days before perihelion at 0.96~AU from the Sun, peaked
1.5~days later,~and involved nearly the entire mass of the comet's original
nucleus (which survived the first outburst).  The mass fragmented into
a cloud of dust grains expanding with a velocity of nearly 80~m~s$^{-1}$.
The cloud could be followed to perihelion, but not beyond.  The dominant
size of the dust in the cloud was determined from the temporal distribution
of residuals, which the debris' center of light exhibited relative to the
gravitational orbit derived from the pre-outburst observations, taken as
offsets of a companion nucleus from the primary nucleus of a split comet.
The residuals were increasing with time from the fragmentation event at
an accelerated rate due to solar radiation pressure whose magnitude
amounted to 0.00216 the Sun's gravitational acceleration, equivalent to
64$\,\times 10^{-8}$~AU~day$^{-2}$ at 1~AU from the Sun and implying a
dominant dust particle diameter of 1~mm at an assumed bulk density of
$\sim$0.5~g~cm$^{-3}$.  The astrometric positions acquired on the last
several days of ground-based observation, 17 to 13~days before perihelion,
were found to deviate steadily from the fit; one observer reported the
comet to be suddenly displaced.  This peculiar effect led to a surprising
conclusion, as noted below.

The derived particle size is significant because it is nearly identical to
the size of the dust that populated the sunward tip of the headless tail of
C/2011~W3 following that comet's disintegration (Sekanina \& Chodas 2012).
Particles in the millimeter-size range are also typical for the debris in
cometary dust trails (e.g., Sykes et al.\ 1990; Reach et al.\ 2007) and for
meteoroids in related meteor outbursts (e.g., Jenniskens 1998); Sykes et
al.\ specifically state that most mass and surface area of cometary trails
are contributed by grains $\sim$1~mm in diameter.  Finally, Fulle et al.\
(2015), describing the results from the GIADA instrument on board the
Rosetta spacecraft, report the detection of {\it charged fluffy aggregates
of submicron-sized grains\/}, whose size range is centered near 1~mm,
whose equivalent bulk density is lower than 0.001~g~cm$^{-3}$, and whose origin is
probably linked to interstellar dust.  Fulle et al.\ add that even though
the aggregates contribute very little to the mass of the dust population
of comet 67P, they account for a substantially higher fraction of the total
cross-sectional area.

\begin{figure*}
\vspace{-2.6cm}
\hspace{-0.21cm}
\centerline{
\scalebox{0.85}{
\includegraphics{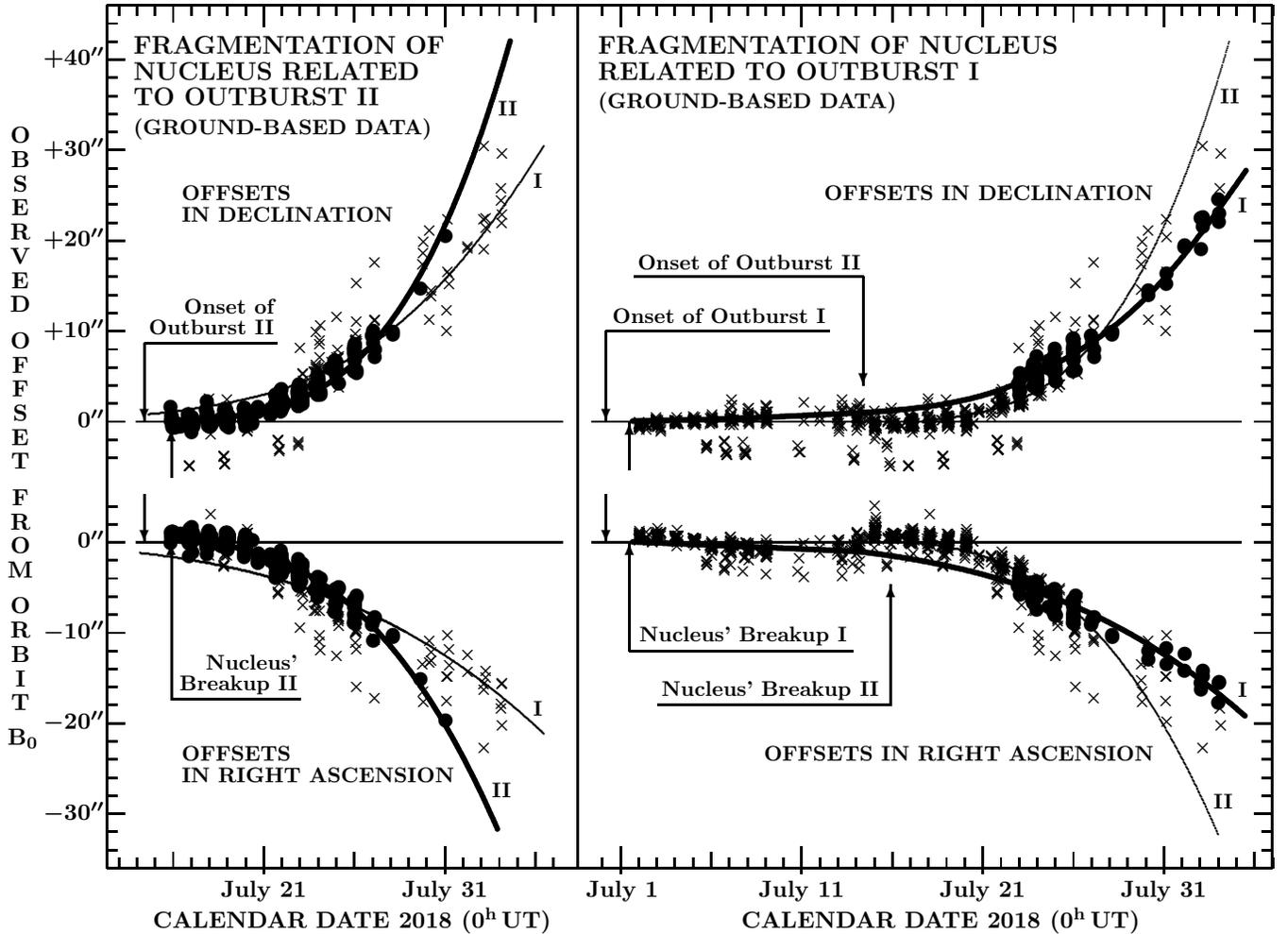}}}
\vspace{-9.52cm}
\caption{Evidence of sudden fragmentation of the nucleus of comet C/2017~S3
associated with the second (left) and first (right) outburst.  The thick
curves fit the residuals, or offsets, from a gravitational orbit (B$_0$) linking
the observations (solid circles) used in the given solution (I for the
debris released in a breakup during the first outburst, II in a breakup during
the second outburst); the crosses, some fitted by the thin curves, are the
ignored observations.  Note that most data points for July~30--August~3, 2018
(17 to 13~days before perihelion) fit the fragment released during the first
outburst, not the dust cloud released during the second outburst. (From
Sekanina \& Kracht 2018.){\vspace{0.45cm}}}
\end{figure*}

The first, nonfatal outburst of comet C/2017~S3 began 46.8~days before perihelion
at 1.25~AU from~the~Sun, peaked about 4~days later, and involved only a
small~fraction of the nucleus' mass (Sekanina \& Kracht 2018).~This event was much
less powerful than the second outburst, and it was doubtful whether a signature
of the debris could be detected.  Surprisingly, however, the
positions secured on the last days of ground-based observation that did not fit
the debris from the second outburst turned out to be consistent with the predicted
positions of the debris from the first outburst.~\mbox{Sekanina}~\&~Kracht~then
encountered a new surprise:\ a crude upper limit on the expansion velocity
{\vspace{-0.04cm}}equaled a fraction of 1~m~s$^{-1}$ (in~contrast to nearly
80~m~s$^{-1}$ for the other dust cloud)~over a period of four weeks, making
it in effect likely~that~(i)~the velocity was zero and (ii)~the ``cloud'' of
dust was a single sizable fragment.  The observations implied the presence of
an antisolar nongravitational acceleration of 0.00057 the Sun's gravitational
acceleration, or 16.9$\,\times 10^{-8}$\,AU~day$^{-2}$ at 1~AU from the Sun.
In the absence of any sign of outgassing, the magnitude of this acceleration
must imply in effect the fragment's very high surface-area-to-mass ratio,
suggesting that it was a monstrous, extremely fluffy aggregate of
loosely-bound dust grains.

This paradigm is further supported by an appropriate nongravitational
orbital solution that linked the pre-outburst observations with the
observations of the bright condensation appearing at the time of the
first outburst.  The nongravitational acceleration thus derived, of
17.4$\,\times 10^{-8}$\,AU~day$^{-2}$ at 1~AU from the Sun, is virtually
identical with the acceleration determined from the data acquired on the
last days of ground-based observation, more than two weeks later.  This
agreement shows that the large fluffy-aggregate fragment was part of the
condensation emerging from the first outburst, whose brightness early on
was made up of gas emissions and, to a much lesser extent, of microscopic
dust ejecta.  Whether the fluffy-aggregate fragment was released in the
course of the first outburst en bloc or reassembled, at submeter-per-second
velocities, from numerous smaller aggregate units shortly after
their nearly simultaneous release is unclear.  In any case, the fragment
was rapidly devolatilized, as documented by the accelerated drop in the
condensation's brightness over the two weeks between the outbursts.  All
debris from the first outburst was subsequently obliterated by the massive
ejecta from the second outburst, showing up as a new condensation, and the
fluffy-aggregate fragment was not detected and bisected for position until
the surface brightness of the new condensation subsided, as a result of
its rapid expansion, to very low levels.  The complex nature of the
distribution of residuals of the center of light measured for orbital
position following the two outbursts is illustrated in Figure~1, copied
from the paper by Sekanina \& Kracht (2018).

\section{Comparison of 1I/`Oumuamua with C/2017 S3}
Bortle's rule thus predicts that, if built fundamentally the same way,
comet C/2017~S3 and 1I should both have disintegrated before perihelion,
displaying only their debris afterwards.  An important difference between
the two objects is that at critical times the former was favorably placed
for observation before perihelion but poorly after perihelion, whereas
it was the other way around~for the latter.  These circumstances contributed
to the failure~to discover 1I before perihelion and to the lack of~information
on the debris of C/2017~S3 after perihelion.

Since 1I was observed only as a single highly irregular object with no
activity and no companions, the most likely counterpart to the presumed
event that the object underwent is, in the C/2017~S3 scenario, the first
outburst with the surviving large fluffy-aggregate fragment.  This would
fit a plausible guess that 1I was far less massive than C/2017~S3.  The
first outburst would consume the entire volume of 1I, leaving no mass
for a second~one.  In addition, one may not rule out a very modest amplitude
of `Oumuamua's outburst, if perceptible at all.

There are other reasons for comparing 1I to C/2017~S3 in general and the
putative outburst of 1I to the first outburst of C/2017~S3 in particular.
One, the best estimates for the flight time of 1I from its home stellar system
to the Solar System range according to Bailer-Jones et al.\ (2018) from 1.0 to
3.8~million years for the four nearby candidates, while the time C/2017~S3
needed to get in from its aphelion in the Oort Cloud 0.5~pc from the Sun is,
 based on the original orbit determined~by~\mbox{Sekanina} \& Kracht (2018),
some 5.7 million years, of which 3~million years was the comet cruising beyond
0.4~pc from the Sun.  Thus, the two objects spent the last several million years
in essentially the same environment.  Two, the nongravitational accelerations
that `Oumuamua and the fluffy-aggregate fragment of C/2017~S3 were subjected
to are similar, 0.00083 and 0.00057 the Sun's gravitational acceleration,
respectively, differing from each other by less than 50~percent and suggesting
that their surface-area-to-mass ratios --- and therefore porosity --- were~alike.
The masses are probably also similar; \mbox{Sekanina} \& Kracht (2018) derived
{\vspace{-0.05cm}}a value of 10$^7$\,g for `Oumuamua, while estimating a crude
upper limit of 1.1$\times 10^9$\,g for the fluffy-aggregate fragment of C/2017~S3.

Given that the nongravitational effect on 1I varied as an inverse square
of heliocentric distance, the verdict on whether it was triggered by an
outgassing-driven~momen\-tum or by solar radiation pressure comes out
unequivocally in favor of the latter.  The failure of the standard Style~II
nongravitational law by Marsden et al.\ (1973) was already pointed out by
Micheli et al.\ (2018), a result that eliminates effects generated by the
\mbox{sublimation}~of water ice and/or more refractory (mostly carbon-based)
substances, but not by the much more volatile~ices,~such as carbon dioxide and
especially carbon monoxide.  Unfortunately, these two ices were ruled out as
potential sources of the orbital perturbation  by the negative~observations
of the Spitzer Space Telescope,~made~at~4.5~$\mu$m~on November \mbox{21--22},
2017 (Trilling et al.\ 2018), when the object was 2.0~AU from the Sun.
The 3$\sigma$ upper limits on the production rate were found to be
9$\,\times 10^{22}$\,molecules~s$^{-1}$ for CO$_2$ and
9$\,\times 10^{21}$\,molecules~s$^{-1}$ for CO.  These rates are equivalent to
rotationally averaged emission surface areas of merely 54~m$^2$ for CO$_2$ and
1.4~m$^2$ for CO, too low by orders of magnitude to explain the nongravitational
effect without involving very high porosity.  These arguments parallel the
objections to the sublimation hypothesis based on the implied rotational
instability (Rafikov 2018).  Needless to say, a {\small \bf dead cometary
fragment does not outgas}.

\section{Comparison of 1I/`Oumuamua with C/2010 X1\\and Possible
 Disintegration Mechanism}
Another extensively investigated comet whose behavior is
potentially relevant to `Oumuamua is C/2010 X1 (Elenin),
probably another Oort Cloud object (Nakano 2011).  Its
perihelion distance was 0.48~AU and absolute magnitude
(at a zero phase angle) \mbox{$H_0 = 11.7$} (Li \& Jewitt
2015).  Since Bortle's survival limit for this object is
9.9, it too was expected to disintegrate before perihelion.
According to Li \& Jewitt, the breakup of the nucleus, which was
at least 1.2~km in diameter, began most probably on August~18,
2011, 23 days before perihelion, at a heliocentric distance
of 0.74~AU.  Discussing disintegration mechanisms, Li \& Jewitt
noted that only a small fraction of the nucleus needs to be peeled
off by sublimation to change the angular momentum by a
large factor, suggesting that rotational breakup by torques
triggered by the outgassing-driven mass loss was a likely candidate.
This idea is similar to that expressed by Rafikov (2018) to
argue against `Oumuamua being an active comet.

Li \& Jewitt (2015) also determined that the extent of
the active surface from which water ice was sublimating
began to drop even before the breakup, suggesting that
the comet's ice reservoir was rapidly being depleted.
On October 22, 2011, 42~days after perihelion, the authors
were looking for the comet's possible debris with the
360-cm CHFT reflector on Mauna Kea, Hawaii.  From the
negative result they concluded that no fragments larger
than 80~meters in diameter had survived.  Similarly,
Kidger et al.\ (2016) reported that no cloud with dust
and/or boulders in the millimeter- to meter-size range
and no discrete fragments larger than 260~meters across
were detected with the Herschel Space Telescope three days
earlier.  On the other hand, Guido et al.\ (2011), using
a 10-cm f/5 wide-field refractor, did detect a headless
tail on October 21 and 23 --- also confirmed by R.\ Ligustri
on October 22 (Dymock 2011) --- whose sharp southern boundary
was interpreted by Sekanina (2011) as a synchrone that
signaled a brief emission event occurring about August~16,
the date essentially coinciding with the onset time of the
comet's disintegration determined by Li \& Jewitt (2015).
He estimated the largest surviving pieces of the nucleus to
be in the centimeter range by the time of observation, with
submillimeter-sized grains at the other end of the size
spectrum that made up the detected tail.

If outgassing-driven rotational breakup is the mechanism
responsible for disintegration of intrinsically faint
comets with small perihelion distances, Li \& Jewitt's (2015)
modeling suggests that the effect varies as the product
of the sublimation rate (i.e., inversely as approximately
the square of heliocentric distance), the rotation period, and
the length of the moment arm of the torque and inversely as the
nucleus' mass.  `Oumuamua's survival in form of a {\small \bf
single sizable aggregate~\mbox{fragment}}~is contemplated as
a result of a {\small \bf ``limited'' disintegration} event,
which implies that the spin-up process fell short of reaching
the ultimate phase:\ it was enough to trigger the tumbling but not
crumbling of the whole mass.  Because the likelihood of
rotational breakup increases with decreasing mass, in order to
maintain `Oumuamua as the product of the limited disintegration
event, it is essential for the parent dwarf comet to have had a
short initial rotation period, a short torque arm, and, primarily,
a very limited repository of ice.  Alternatively, the parent~disintegrated
completely and `Oumuamua was reassembled by coagulation at very low
velocities from parts of the debris shortly after the event.

The proposed extreme degree of porosity of `Oumuamua does not rule out
a degree of cohesion between the aggregate's grains to survive or
even collisionally grow throughout the disintegration event, depending on the
conditions.  Once the parent's activity ceases, rotational fission
of `Oumuamua is not an issue, as pointed out by Trilling
et al.\ (2018).

\section{Conclusions}
In conclusion, `Oumuamua is described as a piece~of debris of a dwarf interstellar
comet, which --- following Bortle's rule for survival of comets in orbits of
small perihelion distance --- had disintegrated at a time not long before
perihelion, possibly because of its nucleus' rotational breakup.  Conceptualized
as a monstrous, very irregularly shaped and {\it devolatilized\/} fluffy aggregate
of loosely-bound dust grains, of extremely high porosity and 10$^7$\,g in mass
(Sekanina \& Kracht 2018), this model of `Oumuamua provides an opportunity to
consistently explain its nongravitational acceleration as an effect~of solar
radiation pressure and to emphasize a chance~that a number of its observed
properties, including the lack of activity, tumbling, radiation pressure
effect,~and~the~inferred appearance, morphology, and elongated shape are of recent
origin, not inherent to the object that was entering the inner Solar System in
early 2017.  The rest of the parent dwarf comet's debris is expected to have
escaped detection after perihelion.  It is pointed out that the unknown timing of
the disintegration event, in the course of which the nongravitational acceleration
began to affect the orbital motion of `Oumuamua,~may~compromise investigations
of the stellar system from which the object had arrived.  The preperihelion
brightness of `Oumuamua's parent remains unknown, but one cannot entirely
rule out the possibility that it was serendipitously detected near, or
during, the putative outburst; limited search for potential images from
August~2017 (when the object was $<$60$^\circ$ from the Sun and could
have been as faint as magnitude 19) in archival records of observatories
is recommended.\\[-0.1cm]

This research was carried out at the Jet Propulsion Laboratory, California
Institute of Technology, under contract with the National Aeronautics and
Space~Administration.\\[-0.35cm]
%
% Steep albedo slope -> dark surface, contrary to Spitzer
%
% \pagebreak
%
\begin{center}
{\footnotesize REFERENCES}
\end{center}
\vspace{-0.4cm}
\begin{description}
{\footnotesize
\item[\hspace{-0.3cm}]
Bailer-Jones, C.\ A.\ L., Farnocchia, D., Meech, K.\ J., et al.\ 2018,{\linebreak}
 {\hspace*{-0.6cm}}AJ, 156, 205% (11pp)
\\[-0.57cm]
\item[\hspace{-0.3cm}]
Bannister, M.\ T., Schwamb, M.\ E., Frazer, W.\ C., et al.\ 2017,{\linebreak}
 {\hspace*{-0.6cm}}ApJL, 851, 38% (7pp)
\\[-0.57cm]
\item[\hspace{-0.3cm}]
Bortle, J.\ E.\ 1991, Int.\ Comet Quart., 13, 89
\\[-0.57cm]
\item[\hspace{-0.3cm}]
Dymock, R.\ 2011, JBAA, 121, 369
\\[-0.57cm]
\item[\hspace{-0.3cm}]
Fulle, M., Della Corte, V., Rotundi, A., et al.\ 2015, ApJL, 802, 12% (5pp)
\\[-0.57cm]

\vspace{0.2cm}
\item[\hspace{-0.3cm}]
Guido, E., Sostero, G., \& Howes, N.\ 2011, CBET 287{6\vspace{0cm}}
\\[-0.57cm]
\item[\hspace{-0.3cm}]
Jenniskens, P.\ 1998, Earth, Plan.\ \& Space, 50, 555
\\[-0.57cm]
%
% \item[\hspace{-0.3cm}]
% Jorda, L., Crovisier, J., \& Green, D.\ W.\ E.\ 1992, in Asteroids,
%   Comets, Meteors 1991, ed. A. W. Harris \& E. Bowell (Houston, TX:\
%   Lunar and Planetary Institute), 285
% \\[-0.57cm]
%
\item[\hspace{-0.3cm}]
Kidger, M.\ R., Altieri, B., M\"{u}ller, T., \& Gracia, J.\ 2016, Earth,{\linebreak}
 {\hspace*{-0.6cm}}Moon \& Plan., 117, 101
\\[-0.57cm]
\item[\hspace{-0.3cm}]
Li, J., \& Jewitt, D.\ 2015, AJ, 149, 133
\\[-0.57cm]
\item[\hspace{-0.3cm}]
Marsden, B.\ G., Sekanina, Z., \& Yeomans, D.\ K.\ 1973, AJ, 78,~211
\\[-0.57cm]
\item[\hspace{-0.3cm}]
Micheli, M., Farnocchia, D., Meech, K.\ J., et al.\ 2018, Nature, 559,{\linebreak}
 {\hspace*{-0.6cm}}223
\\[-0.57cm]
\item[\hspace{-0.3cm}]
Nakano, S.\ 2011, NK 2121
\\[-0.57cm]
\item[\hspace{-0.3cm}]
Rafikov, R.\ R.\ (2018), ApJL, 867, 17
\\[-0.57cm]
\item[\hspace{-0.3cm}]
Reach, W.\ T., Kelley, M.\ S., \& Sykes, M.\ V.\ 2007, Icarus, 191, 298
\\[-0.57cm]
\item[\hspace{-0.3cm}]
Richter, N.\ 1949, Astron.\ Nachr., 277, 12
\\[-0.57cm]
\item[\hspace{-0.3cm}]
Sekanina, Z.\ 1984, Icarus, 58, 81
\\[-0.57cm]
\item[\hspace{-0.3cm}]
Sekanina, Z.\ 2011, CBET 2876
\\[-0.57cm]
\item[\hspace{-0.3cm}]
Sekanina, Z., \& Chodas, P.\ W.\ 2012, ApJ, 757, 127% (33pp)
\\[-0.57cm]
\item[\hspace{-0.3cm}]
Sekanina, Z., \& Kracht, R.\ 2018, eprint arXiv:1812.07054
\\[-0.57cm]
\item[\hspace{-0.3cm}]
Sykes, M.\ V., Lien, D.\ J., \& Walker, R.\ G.\ 1990, Icarus, 86, 236
\\[-0.57cm]
\item[\hspace{-0.3cm}]
Trilling, D.\ E., Mommert, M., Hora, J.\ L., et al.\ 2018, AJ, 156,~261% (9pp)
\\[-0.64cm]
\item[\hspace{-0.3cm}]
Whipple, F.\ L.\ 1978, Moon \& Plan., 18, 343}
\vspace{-0.41cm}
\end{description}
\end{document}